\documentclass[aps,prl,twocolumn,showpacs,superscriptaddress,groupedaddress]{revtex4-1}

\usepackage[utf8x]{inputenc}
\usepackage{graphicx}
\usepackage{dcolumn}
\usepackage{bm}
\usepackage{amsmath}
\usepackage{amssymb}
\usepackage{subfigure}
\usepackage{subfloat}

\begin{document}

\title{Energy Transport in Trapped Ion Chains}

\author{Michael Ramm, Thaned Pruttivarasin, Hartmut H\"affner}

\affiliation{Department of Physics, University of California,  Berkeley, CA 94720, USA}

\email{hhaeffner@berkeley.edu}
\date{\today}

\begin{abstract}
We experimentally study energy transport in chains of trapped ions. We use a pulsed excitation scheme to rapidly add energy to the local motional mode of one of the ions in the chain. Subsequent energy readout allows us to determine how the excitation has propagated throughout the chain. We observe energy revivals that persist for many cycles. We study the behavior with an increasing number of ions of up to 37 in the chain, including a zig-zag configuration. The experimental results agree well with the theory of normal mode evolution. The described system provides an experimental toolbox for the study of thermodynamics of closed systems and energy transport in both classical and quantum regimes. 
\end{abstract}
\maketitle

Energy transport and thermalization in nanoscale systems are of special interest as they pertain to the microscopic origins of statistical mechanics and the functionality of biological systems.  The Fermi-Pasta-Ulam paradox, for instance, involves numerical simulations of a chain of oscillators where it was conjectured that nonlinearity in the potential would give rise to ergodicity and lead the system to eventually thermalize \cite{Fermi1955,Berman2005}. The simulation results prove otherwise and further work on the problem has led to discoveries of soliton solutions in related non-linear systems and the concept of dynamical chaos. In the context of energy transport in biological system, there has been a number of theoretical investigations how the high transport efficiency can arise in a molecule located in a noisy environment at the ambient temperature. The possible explanations include constructive interferences among the possible pathways aided by spatial robustness against decoherence \cite{Scholak2011} and inhibition of destructive interference via dephasing noise \cite{Caruso2009}. 

Chains of trapped ions allow for experimental investigation of energy transport phenomena and oscillator chain models \cite{Pruttivarasin2011}. In contrast to naturally occurring systems, the parameters such as the amount of present non-linearity and decoherence can be tuned precisely with additional optical potentials. Recently, trapped ion chains have been proposed as a model system for study of multidimensional spectroscopic techniques commonly used to probe energy transport in photosynthetic molecules \cite{Gessner2013}. In addition to serving as a model system, ion chains may reveal a number of interesting deviations from expected thermodynamic behavior. When the extreme ions on either side of the chain are coupled to heat baths of different temperatures, one expects a non-linear temperature distribution across the ion chain  \cite{Lin2010}. The non-uniformity of ions trapped in the harmonic potential leads to  non-extensive scaling of thermodynamic quantities and to eigenmodes that differ from phonon-like waves \cite{Morigi2004a, Morigi2004b}.

In this Letter, we present the observed energy transport dynamics in long trapped ion chains. We prepare an out-of-equilibrium state of the chain by rapidly imparting momentum onto a single ion at one end of the chain. We then monitor the energy of the ions in the chain as the initial excitation propagates, leading to multiple revivals of energy. The energy revivals persist for a surprisingly long time indicating that the system does not thermalize on the experimental timescale. Our work extends the results obtained for two ions in the quantum regime \cite{Haze2012a} to much longer chains of up to 37 ions. The resultant dynamics are more complex as they involve participation of a greater number of normal modes of the chain.

In order to observe the energy propagation, both the excitation and the energy readout have to be faster than the coupling between the ions. Intuitively, the dynamics can be understood in terms of the eigenmodes of the ion chain. When a single ion on the end of the chain is excited, the system is not in an eigenstate of the full coupled set of oscillators. Rather, the excitation creates a superposition of the eigenmodes which then evolve at their eigenfrequncies. The time evolution results in the local excitation transferred to other ions in the chain. Rephasing of the participating eigenmodes corresponds to energy revivals. The same reasoning applies in the quantum regime for propagation of a single local phonon observed in \cite{Haze2012a}. 

\begin{figure}
\begin{center}
\includegraphics[width = 0.5 \textwidth]{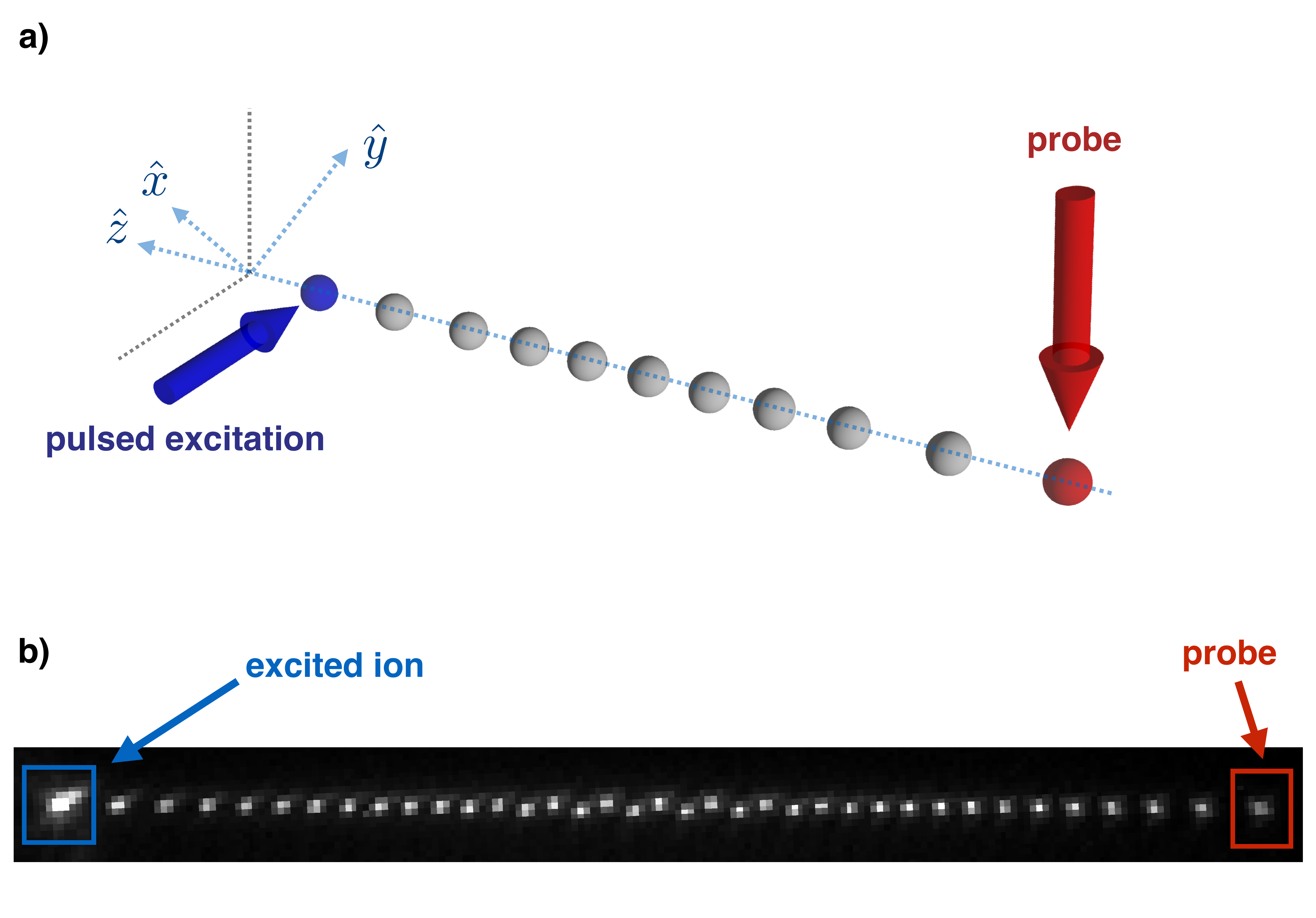}
\caption{\label{schematic} The schematic of the energy transport experiment. (a) The pulsed excitation beam (blue) rapidly adds energy to the ion chain. After a free time evolution, the energy can be read out anywhere along the chain with a probe at 729 nm (red). Both beams are at $45^{\circ}$ with respect to the radial modes of motion.  (b) The CCD image of the ion chain with 37 ions with the tightly focused excitation beam.}
\end{center}
\end{figure}

The experiment proceeds as follows: a chain of $N$ {Ca$^{+}$ ions is confined in a harmonic potential of a linear Paul trap with trap frequencies $\left(\omega_x, \omega_y, \omega_z\right) = 2 \pi \times \left(2.25, 2.0, 0.153\right)  \text{MHz}$. A laser at 397 nm is red-detuned with respect to the S$_{1/2}$-P$_{1/2}$ transition to perform Doppler cooling of the whole chain. An intense beam at 397 nm is tightly focused onto a single ion on one end of the chain, as shown in  Fig.~\ref{schematic}. 

We rapidly add energy using a technique of pulsed excitation: the intensity of a focused beam is switched off and on with the frequency of the local mode \cite{Sheridan2012, Williams2012}}. This resonant process results in a quadratic increase in energy of the excited ion with the number of applied kicks \cite{Pruttivarasin2011}. After 10 $\mu \text{s}$ of pulsed excitation, we stop and let the system evolve freely for a time $\tau$. Then the energy of individual ions in the chain is measured with a laser at 729 nm tuned to the red motional sideband of the narrow quadrupole transition between $| g \rangle = | \text{S}_{1/2};m_J = -1/2 \rangle$ and $ | e \rangle = | \text{D}_{5/2};m_J = -5/2 \rangle$ \cite{Leibfried2003}.

We analyze the dynamics in terms of the eigenmodes of the ion chain by considering that the ions are confined in a potential $V$, which is the sum of the harmonic trap potential and the Coulomb interaction. Minimizing potential energy $V$ with respect to the coordinates of every ion $i$, $\left(x_i,y_i,z_i\right)$, yields the equilibrium positions $\left(x_i^0,y_i^0,z_i^0\right)$. We treat each ion as an individual oscillator coupled to the other ions in the chain. Considering the motion in only one of the radial directions, $x$, the potential energy $V$ can be expanded for small displacements about the equilibrium,  $q_i = x_i - x_i^0$,
\begin{align}
\label{eq:expanded_potential}
V &= \sum_{i=1}^{N}  \frac{1}{2} m \left(\omega_x^2 - \sum_{\substack{j=1 \\ j\neq i}}^{N} \frac{e^2}{4 \pi \epsilon_0 m} \frac{1}{|z_i^0-z_j^0|^3} \right) q_i^2 +\nonumber\\
&+\sum_{i=1}^{N}\sum_{\substack{j=1 \\ j\neq i}}^{N} \frac{1}{2} \frac{e^2}{4 \pi \epsilon_0} \frac{1}{|z_i^0-z_j^0|^3}  q_i q_j
\end{align}
where $e$ is the electron charge and $m$ is the ion mass \cite{James1998,Porras2004}. The local oscillation frequency is modified by the repulsive force from the other ions in the chain and their effect drops off as the cube of the inter-ion distance. The ions are closer to each other in the center of the chain, hence the local oscillation frequency will be minimal for the middle ion. The local displacements $q_i$ are coupled, and the system may be diagonalized in terms of the $N$ radial normal modes of the ion chain \cite{James1998}. We enumerate the normal modes $\vec{v}_n$ in the order of decreasing eigenfrequencies such that $\vec{v}_1$ always refers to the center-of-mass mode with the corresponding eigenfrequency of $\omega_x$.  We model the effect of the pulsed excitation of the leftmost ion as a radial displacement of the ion from its equilibrium position. The unit displacement can decomposed as the sum of radial normal modes with coefficients $c_n$: $\vec{q} = [1, 0, 0, \ldots , 0] = \sum_{i=1}^N c_n \vec{v}_n.$
\begin{figure}
\begin{center}
\includegraphics[width = 0.5 \textwidth]{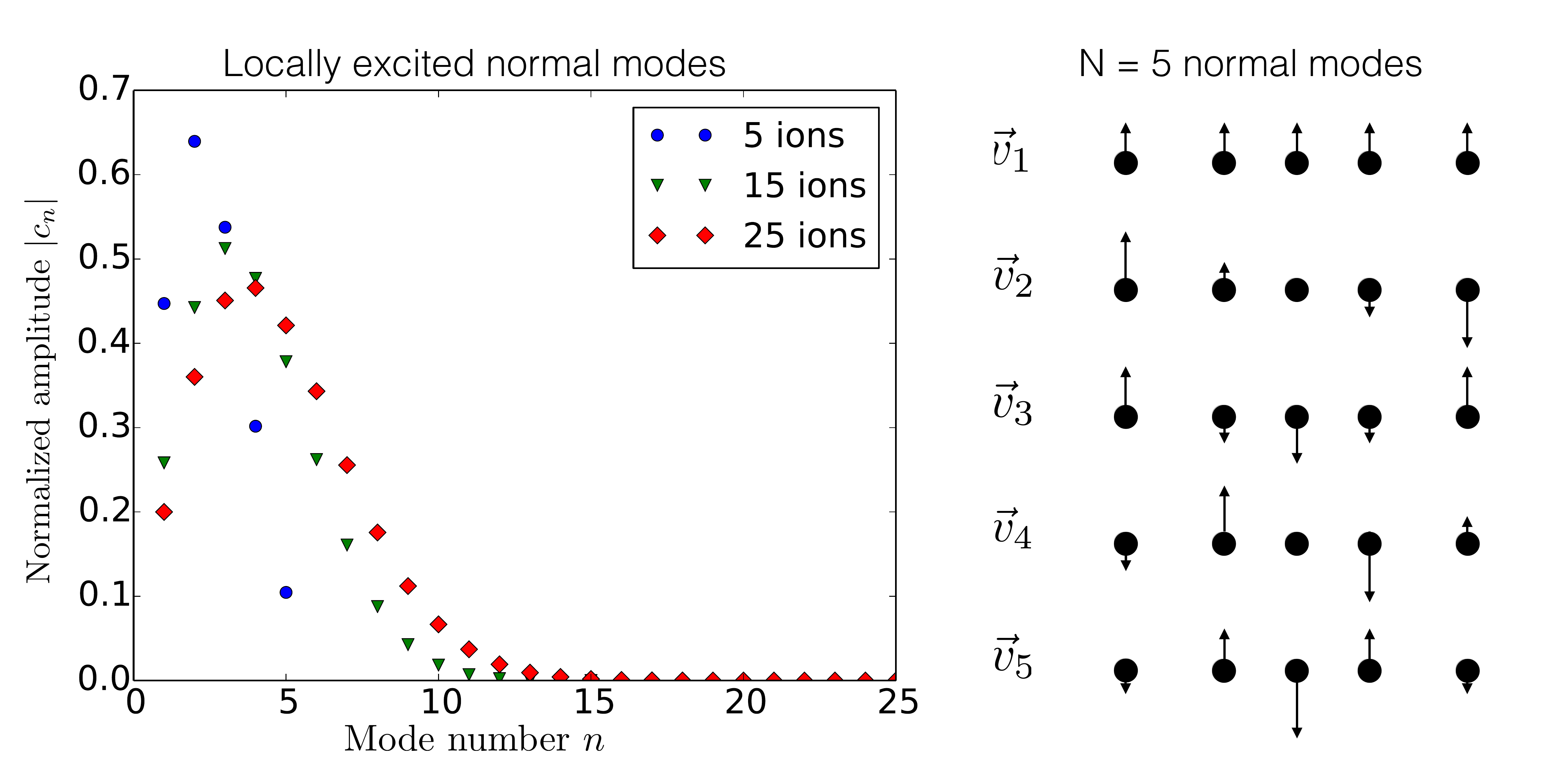}
\caption{\label{normal_modes} Shown on the left is the decomposition of a unit displacement of the leftmost ion in terms of the radial eigenmodes of the ion chain. For longer chains, only lower-ordered modes are excited. On the right are the 5 possible eigenmodes for a chain of $N=5$ ions. Modes $4$ and $5$ are not excited strongly because they do not involve significant motion of the leftmost ion.}
\end{center}
\end{figure}

The eigenmode decomposition is shown in Fig.~\ref{normal_modes} for chains of increasing number of ions. We see that the pulsed excitation creates a superposition of eigenmodes of chain motion, which then evolve at their corresponding eigenfrequencies. However, even for long chains, only the first 10 modes play a large role in determining the subsequent time evolution.

In order to precisely measure the energy of the system after an evolution time $\tau$, we consider the quantized version of the system with the potential energy described by Eq.~(\ref{eq:expanded_potential}) with the Hamiltonian $H_x$ written in terms of local phonons creation and annihilation operators at site $i$ denoted $a_i^\dagger$, $a_i$, respectively \cite{Ivanov2009, Deng2008}:
\begin{align}
{H}_x = \sum_{i=1}^{N} \hbar \omega_{x,i} a_i^\dagger a_i + \hbar \sum_{i=1}^{N} \sum_{\substack{j=1 \\ j < i}}^{N} t_{ij} \left(a_i^\dagger a_j + a_i a_j^\dagger \right)
\end{align}
where we have neglected fast-rotating non-energy conserving terms. The site-dependent oscillation frequency and the tunneling amplitude are given by:
\begin{align}
\omega_{x,i} &= \omega_x - \frac{1}{2} \sum_{\substack{j=1 \\ j \neq i}}^N \frac{1}{m \omega_x} \frac{e^2}{4 \pi \epsilon_0}\frac{1}{|z_i^0-z_j^0|^3}\\
t_{ij} &= \frac{1}{2} \frac{1}{m \omega_x} \frac{e^2}{4 \pi \epsilon_0}\frac{1}{|z_i^0-z_j^0|^3} = \frac{1}{2} \frac{\omega_z^2}{\omega_x} \frac{1}{{\Delta \bar{z}}^3}
\end{align}
where $\Delta \bar{z} \sim O(1)$ is a dimensionless separation of ions in the units of the characteristic length scale \cite{James1998}. We find that for the experimental trap frequencies, the tunneling matrix element between the leftmost ion and its neighbor, $t_{12}$, ranges from $2 \pi \times 6.7 \text{ kHz}$ for a chain of 5 ions to $2 \pi \times 21.1 \text{ kHz}$ for a chain of 25 ions.

After performing Doppler cooling, the local radial mode of every ion is in a thermal state with a temperature $\bar{n}$. The density matrix of the ion $i$, $\rho^{\text{th}}_i$, can be expressed in terms of the local phonon number basis $| n_i \rangle$:
\begin{align}
\rho^{\text{th}}_i = \frac{1}{\bar{n} + 1} \left( \frac{\bar{n}}{\bar{n} + 1} \right)^n | n_i \rangle \langle n_i |
\end{align}
The pulsed excitation process is modeled as a displacement operator $D(\alpha)$ applied onto the leftmost ion ($i=1$) in the chain \cite{Ziesel2013} for some complex amplitude $\alpha = |\alpha|e^{i \phi}$ resulting in a displaced thermal state of motion. Experimentally we do not control the phase of the pulsed laser $\phi$, yielding a diagonal density matrix of the first ion $\rho_1$ after averaging over $\phi$:
\begin{align}
\rho_1 &=  \frac{1}{2\pi} \int d \phi \left( {D}(\alpha) \rho^{\text{th}}_1 {D}{(\alpha)}^\dagger \right) =  \sum_n p_n^{\text{disp}}  | n_1 \rangle \langle n_1 |
\end{align}
with the occupational probabilities given by \cite{Saito1996}:
\begin{equation}
p_n^{\text{disp}} = \left( \frac{1}{\bar{n} + 1} \right) \left( \frac{\bar{n}}{\bar{n} + 1} \right)^{n} e^{-\frac{|\alpha|^2}{\bar{n} + 1}}{L}_{n} \left(-\frac{|\alpha|^2}{\bar{n} (\bar{n} + 1)} \right)
\end{equation}
where $L_n$ is the Laguerre polynomial of the $n^{\text{th}}$ degree. We measure the displacement $|\alpha|$ of the ion $i$ by driving the red motional sideband of the transition between $|g\rangle$, and $| e \rangle$. This interaction couples the electronic and motional states of the ion in the form $|g\rangle | n_i \rangle$ and $|e\rangle | n_i-1 \rangle$ with Rabi frequency $\Omega_{n,n-1}$, which depends on the particular motional state $n$ \cite{Leibfried2003}:
\begin{align}
\Omega_{n,n-1} = \Omega_0 | \langle n - 1 | e^{i \eta \left( a + a^\dagger \right)} | n \rangle | 
\end{align}
where $\Omega_0$ is a scale of the coupling strength and $\eta$ is the Lamb Dicke parameter. In the regime of our experiment, the Rabi frequency $\Omega_{n,n-1}$ increases with the $n$, and the energy can be determined by monitoring the strength of the sideband interaction. Specifically, we measure the probability to find the ion in the electronic ground state $|g\rangle$:
\begin{align}
P_g(t) = \frac{1}{2} \left[1 + \sum_{n=0}^\infty p_n^{\text{disp}} \cos(\Omega_{n,n-1} t)\right]
\end{align}
First, we extract the initial temperature, $\bar{n}$, without pulsed excitation. Then, with pulsed excitation added, the knowledge of $\bar{n}$, the laser intensity, and the trapping parameters allows us to calculate the displacement $|\alpha|$ from the electric ground state probability $P_g(t)$. The experimental excitation time is fixed at $t = 7.5~\mu \text{s}$, short compared to the coupling in order to address only the local mode of motion. The short durations of both the energy readout and the pulsed excitation (10 $\mu \text{s}$) are crucial to observe the energy dynamics as these operations are much faster than the characteristic coupling time of the leftmost ion $2 \pi / t_{12}$.

\begin{figure}
\begin{center}
\includegraphics[width = 0.5 \textwidth]{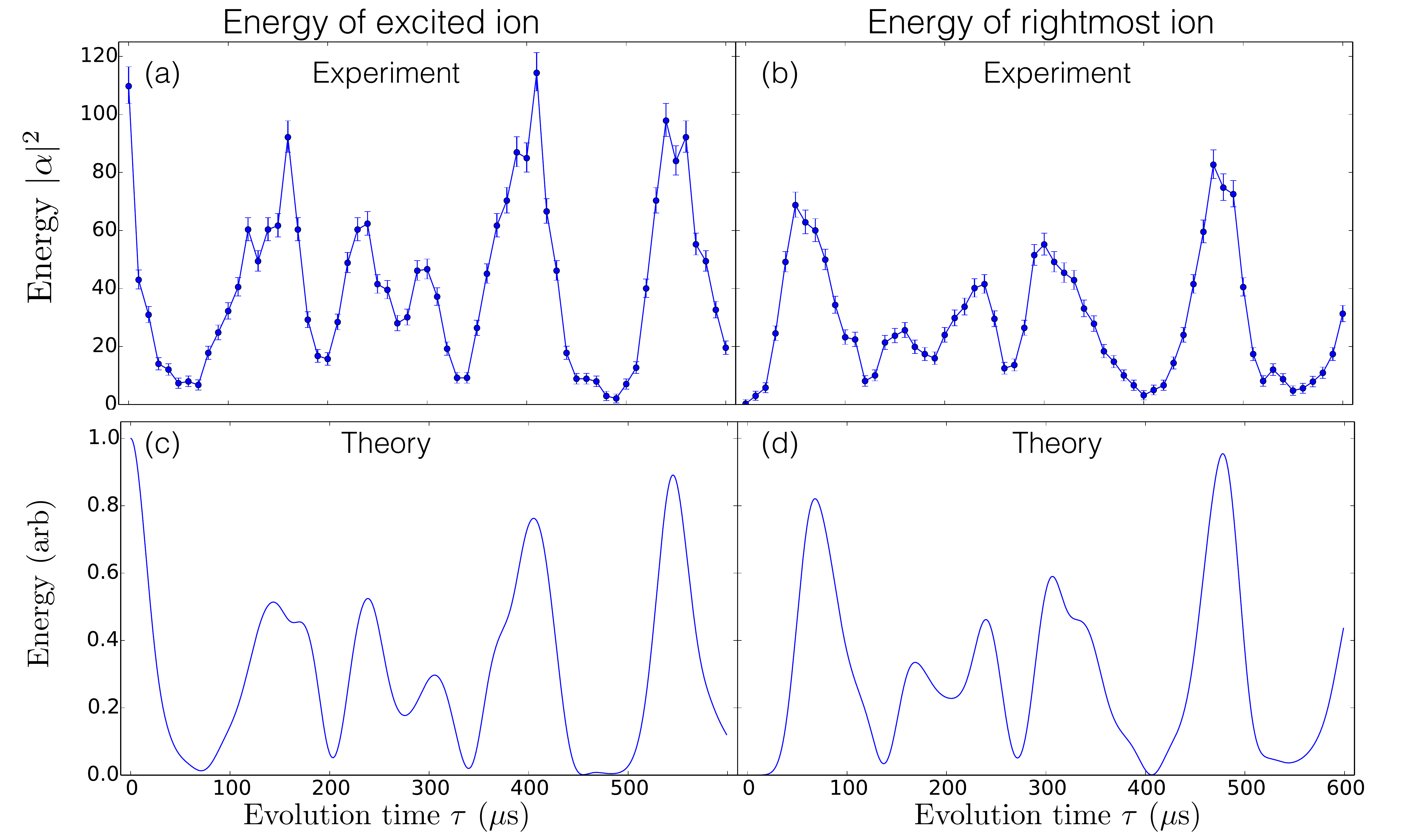}
\caption{\label{theory_comparison_5} Energy transport in a chain of 5 ions. The leftmost ion is given a kick and its energy is measured after a subsequent evolution time $\tau$ as shown in (a). The energy revivals occur at the rephasing times of the eigenfrequencies and approach the energy of the initial excitation.  Plot (b) shows the energy of the rightmost ion: initially unexcited the energy from the kick is rapidly transferred across the chain. The energy of both ions evolves according to the populated normal modes of motion. Each point is a result of 500 measurements resulting in the denoted statistical error bars. The points are connected to guide the eye. Plots (c) and (d) show the kinetic energies of the ions from the molecular dynamics simulations.}
\end{center}
\end{figure}

The results for the 5-ion chain and the comparison to theory are presented in Fig.~\ref{theory_comparison_5}. The experimental data are in good agreement with a molecular dynamics simulation where the leftmost ion is initially displaced in the radial direction. The simulation takes into account the full potential $V$, including non-linearities from the Coulomb repulsion and the driven motion of ions in the Paul trap. It has no free parameters and uses independently measured trap frequencies as inputs. The simulation results show that the dynamics of the time evolution do not change with a decreased initial excitation amplitude, confirming the absence of non-linear effects and justifying the normal mode picture.
\begin{figure}
\begin{center}
\includegraphics[width = 0.5 \textwidth]{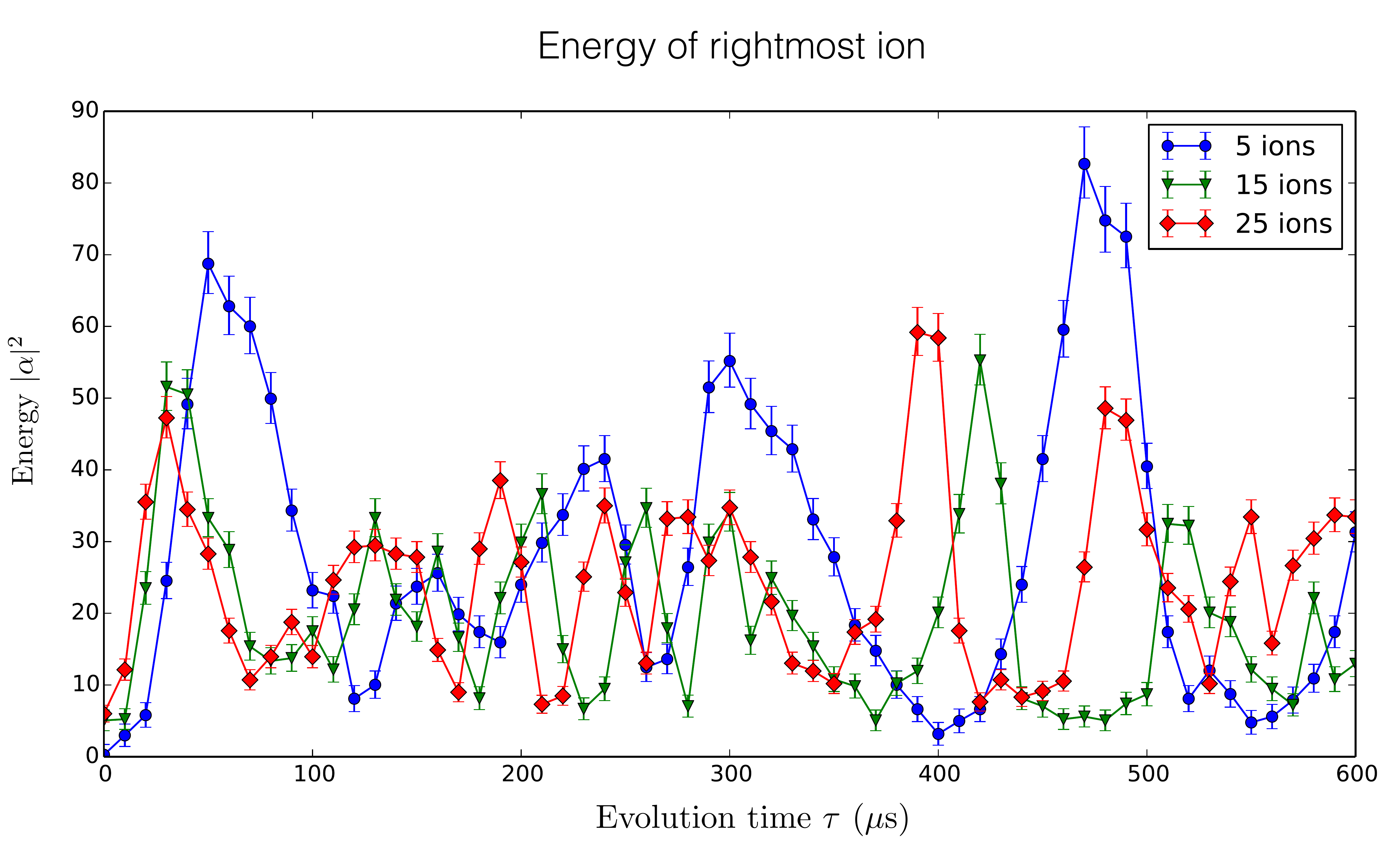}
\caption{\label{right_ion_figure.pdf} The energy of the rightmost ions measured for progressively longer chains. The rate of energy transfer across the chain  increases slightly to the reduction in the inter-ion distance but has a weak dependance on the ion number. Due to the faster coupling, the feature size of the revivals decreases for longer chains. The average measured energy drops: with a greater number of participating normal modes, the excitation is distributed among more ions.}
\end{center}
\end{figure}

We repeat the experiment with progressively longer chains such that a greater number of normal modes is populated by kicking the leftmost ion. The energy of the rightmost ion during the sequence of experiments is presented in Fig.~\ref{right_ion_figure.pdf}. Similar revivals can be identified for increasing chain length. This is explained by the fact that the eigenfrequencies of the populated 10 normal modes have only a weak dependence on the ion number. For example, the splitting between the eigenfrequencies of the modes $\vec{v}_1$ and $\vec{v}_5$ increases by $\sim 3 \%$ as the length is increased from $N=5$ to $N=25$. It can be seen that the revival features become sharper for longer chains: more normal modes participate in the dynamics and the ions are spaced closer together increasing the coupling rate.  The faster coupling also leads to a higher energy of the rightmost ion for evolutions times $\tau \sim 0 ~\mu\text{s}$:  some energy has already transferred from the excited ion to the rightmost ion by the time the pulsed excitation is complete. 

\begin{figure*}
\begin{center}
\includegraphics[width = 1.0 \textwidth]{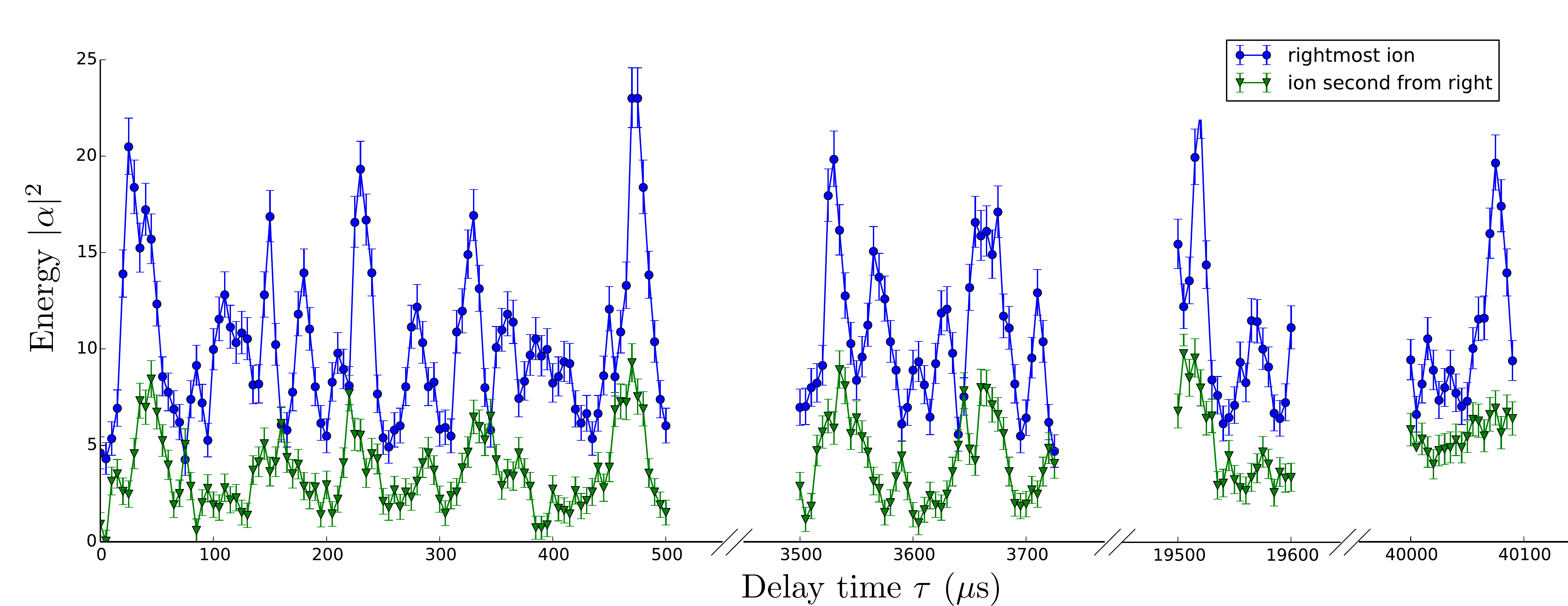}
\caption{\label{together_37} The energy revivals measured for the rightmost ion (circles) and ion second from the right (triangles) in a 37-long ion chain partially in the zig-zag configuration. The dynamics extend up $40~\text{ms}$ and are likely limited by trap frequency instabilities. The energy from the kicked ion does not efficiently transfer to the ion second from the right. The general rise in the measured energy corresponds to background heating during the free evolution.}
\end{center}
\end{figure*}

Even for very long chains, the energy revivals persist for a long time compared to the coupling strength. This is illustrated by Fig.~\ref{together_37}. This measurement was performed with 37 ions in a partially zig-zag configuration as shown in Fig.~\ref{schematic}b. The energy revivals continue even after an evolution time of $\tau = 40~\text{ms}$. For times longer than 40~ms the dynamics wash out, likely due to the instability of the trapping frequencies. Consecutive measurements apart by 12 minutes for the evolution time $\tau = 40~\text{ms}$ revealed a $20~\mu \text{s}$ shift in the position of the revival peak, corresponding to $5 \times 10^{-4}$ change in the coupling strength. The trap frequencies, particularly the radial frequency $\omega_x$, are not expected to be stable at this level.

The measurement presented in Fig.~\ref{together_37} shows a large difference in the excitation between adjacent ions: the rightmost ion in the chain and its neighbor. While the energy efficiently transfers from the leftmost kicked ion to the rightmost ion, the ion second from the right does not get as energetic. This phenomenon follows from the normal mode decomposition of the initial excitation. In the local mode picture, it can also be seen that the efficient transfer of energy occurs because the kicked and the rightmost ion have the same local trap frequency leading to an on-resonant coupling. However, the local frequency of the ion second from the right is different (by approximately the next neighbor coupling),  leading to an off-resonant excitation.

In summary, we have presented experimental results measuring the transport of energy in chains of trapped ions. By exciting the ions and reading out the energy faster than the coupling, we are able to observe the energy propagation from the kicked ion. The dynamics observed for 5 ions agree well with numerical simulations and are explained by the normal mode decomposition of the initial excitation. This work enables the study of the Fermi-Pasta-Ulam problem in both classical and quantum regimes. We plan to investigate how the ion chain thermalizes by engineering additional non-linearities to alter the normal mode structure. Particularly in the quantum regime, it will be possible to use the demonstrated techniques to perform energy transport experiments where the size of the system would render numerical simulations unfeasible. 

\section{Acknowledgment}
We would like to acknowledge useful discussions with M. Gessner. This work was supported by the NSF CAREER Program Grant No. PHY 0955650. M. R. was supported by an award from the U.S. Department of Energy Office of Science Graduate Fellowship Program (DOE SCGF).

\bibliographystyle{aipnum4-1}
\bibliography{bibliography}

%merlin.mbs aipnum4-1.bst 2010-07-25 4.21a (PWD, AO, DPC) hacked
%Control: key (0)
%Control: author (8) initials jnrlst
%Control: editor formatted (1) identically to author
%Control: production of article title (-1) disabled
%Control: page (0) single
%Control: year (1) truncated
%Control: production of eprint (0) enabled
\begin{thebibliography}{19}%
\makeatletter
\providecommand \@ifxundefined [1]{%
 \@ifx{#1\undefined}
}%
\providecommand \@ifnum [1]{%
 \ifnum #1\expandafter \@firstoftwo
 \else \expandafter \@secondoftwo
 \fi
}%
\providecommand \@ifx [1]{%
 \ifx #1\expandafter \@firstoftwo
 \else \expandafter \@secondoftwo
 \fi
}%
\providecommand \natexlab [1]{#1}%
\providecommand \enquote  [1]{``#1''}%
\providecommand \bibnamefont  [1]{#1}%
\providecommand \bibfnamefont [1]{#1}%
\providecommand \citenamefont [1]{#1}%
\providecommand \href@noop [0]{\@secondoftwo}%
\providecommand \href [0]{\begingroup \@sanitize@url \@href}%
\providecommand \@href[1]{\@@startlink{#1}\@@href}%
\providecommand \@@href[1]{\endgroup#1\@@endlink}%
\providecommand \@sanitize@url [0]{\catcode `\\12\catcode `\$12\catcode
  `\&12\catcode `\#12\catcode `\^12\catcode `\_12\catcode `\%12\relax}%
\providecommand \@@startlink[1]{}%
\providecommand \@@endlink[0]{}%
\providecommand \url  [0]{\begingroup\@sanitize@url \@url }%
\providecommand \@url [1]{\endgroup\@href {#1}{\urlprefix }}%
\providecommand \urlprefix  [0]{URL }%
\providecommand \Eprint [0]{\href }%
\providecommand \doibase [0]{http://dx.doi.org/}%
\providecommand \selectlanguage [0]{\@gobble}%
\providecommand \bibinfo  [0]{\@secondoftwo}%
\providecommand \bibfield  [0]{\@secondoftwo}%
\providecommand \translation [1]{[#1]}%
\providecommand \BibitemOpen [0]{}%
\providecommand \bibitemStop [0]{}%
\providecommand \bibitemNoStop [0]{.\EOS\space}%
\providecommand \EOS [0]{\spacefactor3000\relax}%
\providecommand \BibitemShut  [1]{\csname bibitem#1\endcsname}%
\let\auto@bib@innerbib\@empty
%</preamble>
\bibitem [{\citenamefont {Fermi}, \citenamefont {Pasta},\ and\ \citenamefont
  {Ulam}(1955)}]{Fermi1955}%
  \BibitemOpen
  \bibfield  {author} {\bibinfo {author} {\bibfnamefont {E.}~\bibnamefont
  {Fermi}}, \bibinfo {author} {\bibfnamefont {J.}~\bibnamefont {Pasta}}, \ and\
  \bibinfo {author} {\bibfnamefont {S.}~\bibnamefont {Ulam}},\ }\href
  {http://www.osti.gov/accomplishments/pdf/A80037041/01.pdf} {\bibfield
  {journal} {\bibinfo  {journal} {Los Alamos Report LA-1940}\ } (\bibinfo
  {year} {1955})}\BibitemShut {NoStop}%
\bibitem [{\citenamefont {Berman}\ and\ \citenamefont
  {Izrailev}(2005)}]{Berman2005}%
  \BibitemOpen
  \bibfield  {author} {\bibinfo {author} {\bibfnamefont {G.~P.}\ \bibnamefont
  {Berman}}\ and\ \bibinfo {author} {\bibfnamefont {F.~M.}\ \bibnamefont
  {Izrailev}},\ }\href {\doibase http://dx.doi.org/10.1063/1.1855036}
  {\bibfield  {journal} {\bibinfo  {journal} {Chaos}\ }\textbf {\bibinfo
  {volume} {15}},\ \bibinfo {eid} {015104} (\bibinfo {year}
  {2005})}\BibitemShut {NoStop}%
\bibitem [{\citenamefont {Scholak}\ \emph {et~al.}(2011)\citenamefont
  {Scholak}, \citenamefont {de~Melo}, \citenamefont {Wellens}, \citenamefont
  {Mintert},\ and\ \citenamefont {Buchleitner}}]{Scholak2011}%
  \BibitemOpen
  \bibfield  {author} {\bibinfo {author} {\bibfnamefont {T.}~\bibnamefont
  {Scholak}}, \bibinfo {author} {\bibfnamefont {F.}~\bibnamefont {de~Melo}},
  \bibinfo {author} {\bibfnamefont {T.}~\bibnamefont {Wellens}}, \bibinfo
  {author} {\bibfnamefont {F.}~\bibnamefont {Mintert}}, \ and\ \bibinfo
  {author} {\bibfnamefont {A.}~\bibnamefont {Buchleitner}},\ }\href {\doibase
  10.1103/PhysRevE.83.021912} {\bibfield  {journal} {\bibinfo  {journal} {Phys.
  Rev. E}\ }\textbf {\bibinfo {volume} {83}},\ \bibinfo {pages} {021912}
  (\bibinfo {year} {2011})}\BibitemShut {NoStop}%
\bibitem [{\citenamefont {Caruso}\ \emph {et~al.}(2009)\citenamefont {Caruso},
  \citenamefont {Chin}, \citenamefont {Datta}, \citenamefont {Huelga},\ and\
  \citenamefont {Plenio}}]{Caruso2009}%
  \BibitemOpen
  \bibfield  {author} {\bibinfo {author} {\bibfnamefont {F.}~\bibnamefont
  {Caruso}}, \bibinfo {author} {\bibfnamefont {A.~W.}\ \bibnamefont {Chin}},
  \bibinfo {author} {\bibfnamefont {A.}~\bibnamefont {Datta}}, \bibinfo
  {author} {\bibfnamefont {S.~F.}\ \bibnamefont {Huelga}}, \ and\ \bibinfo
  {author} {\bibfnamefont {M.~B.}\ \bibnamefont {Plenio}},\ }\href {\doibase
  http://dx.doi.org/10.1063/1.3223548} {\bibfield  {journal} {\bibinfo
  {journal} {J. Chem. Phys.}\ }\textbf {\bibinfo {volume} {131}},\ \bibinfo
  {eid} {105106} (\bibinfo {year} {2009})}\BibitemShut {NoStop}%
\bibitem [{\citenamefont {Pruttivarasin}\ \emph {et~al.}(2011)\citenamefont
  {Pruttivarasin}, \citenamefont {Ramm}, \citenamefont {Talukdar},
  \citenamefont {Kreuter},\ and\ \citenamefont
  {H\"{a}ffner}}]{Pruttivarasin2011}%
  \BibitemOpen
  \bibfield  {author} {\bibinfo {author} {\bibfnamefont {T.}~\bibnamefont
  {Pruttivarasin}}, \bibinfo {author} {\bibfnamefont {M.}~\bibnamefont {Ramm}},
  \bibinfo {author} {\bibfnamefont {I.}~\bibnamefont {Talukdar}}, \bibinfo
  {author} {\bibfnamefont {A.}~\bibnamefont {Kreuter}}, \ and\ \bibinfo
  {author} {\bibfnamefont {H.}~\bibnamefont {H\"{a}ffner}},\ }\href {\doibase
  10.1088/1367-2630/13/7/075012} {\bibfield  {journal} {\bibinfo  {journal}
  {New J. Phys.}\ }\textbf {\bibinfo {volume} {13}},\ \bibinfo {pages} {075012}
  (\bibinfo {year} {2011})}\BibitemShut {NoStop}%
\bibitem [{\citenamefont {{Gessner}}\ \emph {et~al.}()\citenamefont
  {{Gessner}}, \citenamefont {{Schlawin}}, \citenamefont {{Haeffner}},
  \citenamefont {{Mukamel}},\ and\ \citenamefont
  {{Buchleitner}}}]{Gessner2013}%
  \BibitemOpen
  \bibfield  {author} {\bibinfo {author} {\bibfnamefont {M.}~\bibnamefont
  {{Gessner}}}, \bibinfo {author} {\bibfnamefont {F.}~\bibnamefont
  {{Schlawin}}}, \bibinfo {author} {\bibfnamefont {H.}~\bibnamefont
  {{Haeffner}}}, \bibinfo {author} {\bibfnamefont {S.}~\bibnamefont
  {{Mukamel}}}, \ and\ \bibinfo {author} {\bibfnamefont {A.}~\bibnamefont
  {{Buchleitner}}},\ }\href@noop {} {\ }\Eprint
  {http://arxiv.org/abs/1312.3365} {arXiv:1312.3365} \BibitemShut {NoStop}%
\bibitem [{\citenamefont {Lin}\ and\ \citenamefont {Duan}(2011)}]{Lin2010}%
  \BibitemOpen
  \bibfield  {author} {\bibinfo {author} {\bibfnamefont {G.-D.}\ \bibnamefont
  {Lin}}\ and\ \bibinfo {author} {\bibfnamefont {L.-M.}\ \bibnamefont {Duan}},\
  }\href {\doibase 10.1088/1367-2630/13/7/075015} {\bibfield  {journal}
  {\bibinfo  {journal} {New J. Phys.}\ }\textbf {\bibinfo {volume} {13}},\
  \bibinfo {pages} {075015} (\bibinfo {year} {2011})}\BibitemShut {NoStop}%
\bibitem [{\citenamefont {Morigi}\ and\ \citenamefont
  {Fishman}(2004{\natexlab{a}})}]{Morigi2004a}%
  \BibitemOpen
  \bibfield  {author} {\bibinfo {author} {\bibfnamefont {G.}~\bibnamefont
  {Morigi}}\ and\ \bibinfo {author} {\bibfnamefont {S.}~\bibnamefont
  {Fishman}},\ }\href {\doibase 10.1103/PhysRevLett.93.170602} {\bibfield
  {journal} {\bibinfo  {journal} {Phys. Rev. Lett.}\ }\textbf {\bibinfo
  {volume} {93}},\ \bibinfo {pages} {170602} (\bibinfo {year}
  {2004}{\natexlab{a}})}\BibitemShut {NoStop}%
\bibitem [{\citenamefont {Morigi}\ and\ \citenamefont
  {Fishman}(2004{\natexlab{b}})}]{Morigi2004b}%
  \BibitemOpen
  \bibfield  {author} {\bibinfo {author} {\bibfnamefont {G.}~\bibnamefont
  {Morigi}}\ and\ \bibinfo {author} {\bibfnamefont {S.}~\bibnamefont
  {Fishman}},\ }\href {\doibase 10.1103/PhysRevE.70.066141} {\bibfield
  {journal} {\bibinfo  {journal} {Phys. Rev. E}\ }\textbf {\bibinfo {volume}
  {70}},\ \bibinfo {pages} {066141} (\bibinfo {year}
  {2004}{\natexlab{b}})}\BibitemShut {NoStop}%
\bibitem [{\citenamefont {Haze}\ \emph {et~al.}(2012)\citenamefont {Haze},
  \citenamefont {Tateishi}, \citenamefont {Noguchi}, \citenamefont {Toyoda},\
  and\ \citenamefont {Urabe}}]{Haze2012a}%
  \BibitemOpen
  \bibfield  {author} {\bibinfo {author} {\bibfnamefont {S.}~\bibnamefont
  {Haze}}, \bibinfo {author} {\bibfnamefont {Y.}~\bibnamefont {Tateishi}},
  \bibinfo {author} {\bibfnamefont {A.}~\bibnamefont {Noguchi}}, \bibinfo
  {author} {\bibfnamefont {K.}~\bibnamefont {Toyoda}}, \ and\ \bibinfo {author}
  {\bibfnamefont {S.}~\bibnamefont {Urabe}},\ }\href {\doibase
  10.1103/PhysRevA.85.031401} {\bibfield  {journal} {\bibinfo  {journal} {Phys.
  Rev. A}\ }\textbf {\bibinfo {volume} {85}},\ \bibinfo {pages} {031401}
  (\bibinfo {year} {2012})}\BibitemShut {NoStop}%
\bibitem [{\citenamefont {Sheridan}\ \emph {et~al.}(2012)\citenamefont
  {Sheridan}, \citenamefont {Seymour-Smith}, \citenamefont {Gardner},\ and\
  \citenamefont {Keller}}]{Sheridan2012}%
  \BibitemOpen
  \bibfield  {author} {\bibinfo {author} {\bibfnamefont {K.}~\bibnamefont
  {Sheridan}}, \bibinfo {author} {\bibfnamefont {N.}~\bibnamefont
  {Seymour-Smith}}, \bibinfo {author} {\bibfnamefont {A.}~\bibnamefont
  {Gardner}}, \ and\ \bibinfo {author} {\bibfnamefont {M.}~\bibnamefont
  {Keller}},\ }\href {\doibase 10.1140/epjd/e2012-30377-8} {\bibfield
  {journal} {\bibinfo  {journal} {Eur. Phys. J. D}\ }\textbf {\bibinfo {volume}
  {66}},\ \bibinfo {pages} {289} (\bibinfo {year} {2012})}\BibitemShut
  {NoStop}%
\bibitem [{\citenamefont {Lin}, \citenamefont {Williams},\ and\ \citenamefont
  {Odom}(2013)}]{Williams2012}%
  \BibitemOpen
  \bibfield  {author} {\bibinfo {author} {\bibfnamefont {Y.-W.}\ \bibnamefont
  {Lin}}, \bibinfo {author} {\bibfnamefont {S.}~\bibnamefont {Williams}}, \
  and\ \bibinfo {author} {\bibfnamefont {B.~C.}\ \bibnamefont {Odom}},\ }\href
  {\doibase 10.1103/PhysRevA.87.011402} {\bibfield  {journal} {\bibinfo
  {journal} {Phys. Rev. A}\ }\textbf {\bibinfo {volume} {87}},\ \bibinfo
  {pages} {011402} (\bibinfo {year} {2013})}\BibitemShut {NoStop}%
\bibitem [{\citenamefont {Leibfried}\ \emph {et~al.}(2003)\citenamefont
  {Leibfried}, \citenamefont {Blatt}, \citenamefont {Monroe},\ and\
  \citenamefont {Wineland}}]{Leibfried2003}%
  \BibitemOpen
  \bibfield  {author} {\bibinfo {author} {\bibfnamefont {D.}~\bibnamefont
  {Leibfried}}, \bibinfo {author} {\bibfnamefont {R.}~\bibnamefont {Blatt}},
  \bibinfo {author} {\bibfnamefont {C.}~\bibnamefont {Monroe}}, \ and\ \bibinfo
  {author} {\bibfnamefont {D.}~\bibnamefont {Wineland}},\ }\href
  {http://link.aps.org/doi/10.1103/RevModPhys.75.281} {\bibfield  {journal}
  {\bibinfo  {journal} {Rev. Mod. Phys.}\ }\textbf {\bibinfo {volume} {75}},\
  \bibinfo {pages} {281} (\bibinfo {year} {2003})}\BibitemShut {NoStop}%
\bibitem [{\citenamefont {James}(1998)}]{James1998}%
  \BibitemOpen
  \bibfield  {author} {\bibinfo {author} {\bibfnamefont {D.}~\bibnamefont
  {James}},\ }\href {\doibase 10.1007/s003400050373} {\bibfield  {journal}
  {\bibinfo  {journal} {Appl. Phys. B}\ }\textbf {\bibinfo {volume} {66}},\
  \bibinfo {pages} {181} (\bibinfo {year} {1998})}\BibitemShut {NoStop}%
\bibitem [{\citenamefont {Porras}\ and\ \citenamefont
  {Cirac}(2004)}]{Porras2004}%
  \BibitemOpen
  \bibfield  {author} {\bibinfo {author} {\bibfnamefont {D.}~\bibnamefont
  {Porras}}\ and\ \bibinfo {author} {\bibfnamefont {J.~I.}\ \bibnamefont
  {Cirac}},\ }\href {\doibase 10.1103/PhysRevLett.92.207901} {\bibfield
  {journal} {\bibinfo  {journal} {Phys. Rev. Lett.}\ }\textbf {\bibinfo
  {volume} {92}},\ \bibinfo {pages} {207901} (\bibinfo {year}
  {2004})}\BibitemShut {NoStop}%
\bibitem [{\citenamefont {Ivanov}\ \emph {et~al.}(2009)\citenamefont {Ivanov},
  \citenamefont {Ivanov}, \citenamefont {Vitanov}, \citenamefont {Mering},
  \citenamefont {Fleischhauer},\ and\ \citenamefont {Singer}}]{Ivanov2009}%
  \BibitemOpen
  \bibfield  {author} {\bibinfo {author} {\bibfnamefont {P.~A.}\ \bibnamefont
  {Ivanov}}, \bibinfo {author} {\bibfnamefont {S.~S.}\ \bibnamefont {Ivanov}},
  \bibinfo {author} {\bibfnamefont {N.~V.}\ \bibnamefont {Vitanov}}, \bibinfo
  {author} {\bibfnamefont {A.}~\bibnamefont {Mering}}, \bibinfo {author}
  {\bibfnamefont {M.}~\bibnamefont {Fleischhauer}}, \ and\ \bibinfo {author}
  {\bibfnamefont {K.}~\bibnamefont {Singer}},\ }\href {\doibase
  10.1103/PhysRevA.80.060301} {\bibfield  {journal} {\bibinfo  {journal} {Phys.
  Rev. A}\ }\textbf {\bibinfo {volume} {80}},\ \bibinfo {pages} {060301}
  (\bibinfo {year} {2009})}\BibitemShut {NoStop}%
\bibitem [{\citenamefont {Deng}, \citenamefont {Porras},\ and\ \citenamefont
  {Cirac}(2008)}]{Deng2008}%
  \BibitemOpen
  \bibfield  {author} {\bibinfo {author} {\bibfnamefont {X.-L.}\ \bibnamefont
  {Deng}}, \bibinfo {author} {\bibfnamefont {D.}~\bibnamefont {Porras}}, \ and\
  \bibinfo {author} {\bibfnamefont {J.~I.}\ \bibnamefont {Cirac}},\ }\href
  {\doibase 10.1103/PhysRevA.77.033403} {\bibfield  {journal} {\bibinfo
  {journal} {Phys. Rev. A}\ }\textbf {\bibinfo {volume} {77}},\ \bibinfo
  {pages} {033403} (\bibinfo {year} {2008})}\BibitemShut {NoStop}%
\bibitem [{\citenamefont {Ziesel}\ \emph {et~al.}(2013)\citenamefont {Ziesel},
  \citenamefont {Ruster}, \citenamefont {Walther}, \citenamefont {Kaufmann},
  \citenamefont {Dawkins}, \citenamefont {Singer}, \citenamefont
  {Schmidt-Kaler},\ and\ \citenamefont {Poschinger}}]{Ziesel2013}%
  \BibitemOpen
  \bibfield  {author} {\bibinfo {author} {\bibfnamefont {F.}~\bibnamefont
  {Ziesel}}, \bibinfo {author} {\bibfnamefont {T.}~\bibnamefont {Ruster}},
  \bibinfo {author} {\bibfnamefont {A.}~\bibnamefont {Walther}}, \bibinfo
  {author} {\bibfnamefont {H.}~\bibnamefont {Kaufmann}}, \bibinfo {author}
  {\bibfnamefont {S.}~\bibnamefont {Dawkins}}, \bibinfo {author} {\bibfnamefont
  {K.}~\bibnamefont {Singer}}, \bibinfo {author} {\bibfnamefont
  {F.}~\bibnamefont {Schmidt-Kaler}}, \ and\ \bibinfo {author} {\bibfnamefont
  {U.~G.}\ \bibnamefont {Poschinger}},\ }\href {\doibase
  10.1088/0953-4075/46/10/104008} {\bibfield  {journal} {\bibinfo  {journal}
  {J. Phys. B: At. Mol. Opt. Phys.}\ }\textbf {\bibinfo {volume} {46}},\
  \bibinfo {pages} {104008} (\bibinfo {year} {2013})}\BibitemShut {NoStop}%
\bibitem [{\citenamefont {Saito}\ and\ \citenamefont
  {Hyuga}(1996)}]{Saito1996}%
  \BibitemOpen
  \bibfield  {author} {\bibinfo {author} {\bibfnamefont {H.}~\bibnamefont
  {Saito}}\ and\ \bibinfo {author} {\bibfnamefont {H.}~\bibnamefont {Hyuga}},\
  }\href {\doibase 10.1143/JPSJ.65.1648} {\bibfield  {journal} {\bibinfo
  {journal} {J. Phys. Soc. Jpn.}\ }\textbf {\bibinfo {volume} {65}},\ \bibinfo
  {pages} {1648} (\bibinfo {year} {1996})}\BibitemShut {NoStop}%
\end{thebibliography}%

\end{document}